\documentclass{sf2a-conf2017}

\usepackage{graphicx}
\usepackage{physics}
\usepackage{hyperref}
\usepackage[]{natbib}  
\usepackage{epstopdf}

\def\BibTeX{{\rm B\kern-.05em{\sc i\kern-.025em b}\kern-.08em
    T\kern-.1667em\lower.7ex\hbox{E}\kern-.125emX}}
\bibpunct{(}{)}{;}{a}{}{,}  

\usepackage{amsmath,amssymb}
\usepackage{upgreek}
\usepackage{subcaption}
\usepackage{bm}

\usepackage{ulem}

\usepackage{pgf, tikz}
\usepackage{tikz-3dplot}
\usepackage{pgfplots}
\usetikzlibrary{calc}
\usetikzlibrary{shapes.misc}
\usetikzlibrary{shapes,snakes}
\usetikzlibrary{arrows}
\usetikzlibrary{automata,positioning}
\tikzset{
  load/.style   = {ultra thick,-latex},
  stress/.style = {-latex},
  dim/.style    = {latex-latex},
  axis/.style   = {-latex},
}
\tikzset{dimetric2/.style={
  x={(0.935cm,-0.118cm)},
  y={(0.354cm, 0.312cm)},
  z={(0.000cm, 0.943cm)},
}}
\usepackage{color}
\definecolor{auburn}{rgb}{0.43, 0.21, 0.1}
\definecolor{oxfordblue}{rgb}{0.0, 0.13, 0.28}
\definecolor{coolblack}{rgb}{0.0, 0.18, 0.39}
\definecolor{darkcerulean}{rgb}{0.03, 0.27, 0.49}
\definecolor{denim}{rgb}{0.08, 0.38, 0.74}
\definecolor{purpletaupe}{rgb}{0.31, 0.25, 0.3}
\definecolor{purpletaupemodif}{rgb}{0, 0, 0}
\definecolor{applegreen}{rgb}{0.55, 0.71, 0.0}
\definecolor{chaptercolor}{rgb}{0.2, 0.2, 0.2}
\definecolor{blue(ryb)}{rgb}{0.01, 0.28, 1.0}
\definecolor{ao}{rgb}{0.0, 0.0, 1.0}
\definecolor{electricultramarine}{rgb}{0.25, 0.0, 0.95}
\definecolor{electricpurple}{rgb}{0.75, 0.0, 1.0}
\definecolor{palatinatepurple}{rgb}{0.41, 0.16, 0.38}
\definecolor{aureolin}{rgb}{0.99, 0.93, 0.0}
\definecolor{fluorescentorange}{rgb}{1.0, 0.85, 0.0} 

\defcitealias{ABM2017}{ABM17}

\begin{document}

\TitreGlobal{SF2A 2017}


\title{Layered semi-convection and tides in giant planet interiors}

\runningtitle{Layered semi-convection and tides}

\author{Q. Andr\'e}\address{Laboratoire AIM Paris-Saclay, CEA/DRF -- CNRS -- Universit\'e Paris-Diderot, IRFU/DAp Centre de Saclay, 91191 Gif-sur-Yvette, France}

\author{S. Mathis$^{1,}$}\address{LESIA, Observatoire de Paris, PSL Research University, CNRS, Sorbonne Universités, UPMC Univ. Paris 06, Univ. Paris Diderot, Sorbonne Paris Cité, 5 place Jules Janssen, F-92195 Meudon, France}



\author{A. J. Barker}\address{Department of Applied Mathematics, School of Mathematics, University of Leeds, Leeds, LS2 9JT, UK}

\setcounter{page}{237}


\maketitle


\begin{abstract}
Layered semi-convection could operate in giant planets, potentially explaining the constraints on the heavy elements distribution in Jupiter deduced recently from \textsc{juno} observations, and contributing to Saturn's luminosity excess or the abnormally large radius of some hot Jupiters.
{This is a state consisting of} density staircases, in which convective layers are separated by thin stably stratified interfaces. 
The efficiency of tidal dissipation in a planet depends {strongly} on its internal structure. It is crucial to improve our understanding of the mechanisms driving this dissipation, {since it has important consequences to predict} the long-term evolution of any planetary system. In this work, our goal is to study the resulting tidal dissipation when internal waves are excited by other bodies (such as {the moons of giant planets}) in a region of layered semi-convection. We find that the rates of tidal dissipation can be significantly enhanced in a layered semi-convective medium compared to a uniformly convective medium, especially {in the astrophysically relevant sub-inertial frequency range}. Thus, layered semi-convection is a possible candidate to explain high tidal dissipation rates recently observed in Jupiter and Saturn.
\end{abstract}



\section{Context}
Based on astrometric measurements spanning more than a century \cite{LaineyEtal2009, LaineyEtal2012, LaineyEtal2017} found that the rates of tidal dissipation in Jupiter and Saturn {are} higher than previously thought. This has important astrophysical consequences since tidal interactions {are} a key mechanism {for driving} the rotational, orbital and thermal evolution of moons, planets and stars over very long time-scales.

Moreover, we know that this evolution, linked to the efficiency of tidal dissipation in celestial bodies, strongly depends on their internal structure{s}, {which are} poorly constrained \citep[we refer the reader to the reviews][]{MathisRemus2013,Ogilvie2014}. Several recent studies have suggested new models of giant planet interiors that significantly depart from the standard three-layers model in which a molecular H/He envelope surrounds a metallic H/He envelope, on top of a core composed of heavy elements. Among these works, some suggest that there could be regions exhibiting a stable compositional gradient, either at the core boundary due to {erosion of the core} \citep{GuillotEtal2004,MazevetEtal2015}, or at the interface between metallic and molecular H/He due to settling of He droplets in the molecular region \citep{StevensonSalpeter1977,NettelmannEtal2015}. These predictions seem to be corroborated by recent observations by the \textsc{juno} spacecraft, which  are consistent with interior models of Jupiter {in} which the heavy elements of the core are diluted in the envelope \citep{WahlEtal2017}. Then, the presence of a stabilising compositional gradient alongside the destabilising entropy gradient (driving the convective instability) could lead to layered semi-convection \citep{LeconteChabrier2012,WoodEtal2013}, in which a large number of convective layers are separated by thin stably stratified interfaces. The associated density profile is nearly constant in the convective steps and undergoes a nearly discontinuous jump in stably stratified interfaces, giving a density staircase-like structure. Such structures are observed on Earth, for instance in the Arctic Ocean \citep{GhaemsaidiEtal2016}.

{In} this work, {we} study the impact of layered semi-convection upon the efficiency of tidal dissipation. {T}wo tidal components are usually distinguished: the {equilibrium tide}{,} a large-scale flow induced by the hydrostatic adjustment to the gravitational potential of the perturber \citep[such as the moons of giant planets, see][]{Zahn1966,RemusMathisZahn2012}, {and} the {dynamical tide}, composed of internal waves excited by the perturber \citep{Zahn1975,OgilvieLin2004}. In this context, {the latter} can be called {tidal waves}. Their dissipation by viscosity and thermal diffusion will lead to the long-term rotational, orbital and thermal evolution of the system. This study focuses on how the dynamical tide is affected by the presence of layered semi-convection.

\section{Statement of the problem}
\begin{figure}
\begin{subfigure}{0.38\textwidth}
\centering
\begin{tikzpicture}[scale=0.42]
\pgfmathsetmacro{\cubex}{5}
\pgfmathsetmacro{\cubey}{8}
\pgfmathsetmacro{\cubez}{5}
\pgfmathsetmacro{\centerx}{-2.4*\cubex}
\pgfmathsetmacro{\centery}{-\cubey/2}
\pgfmathsetmacro{\centerz}{-\cubez/2}
\pgfmathsetmacro{\planetradius}{1.1*\cubex/2}
%
\shadedraw[shading=radial,outer color=fluorescentorange!20,inner color=fluorescentorange,draw=none] (\centerx,\centery,\centerz) circle (\planetradius);
\fill[color=red!60] (\centerx,\centery,\centerz) circle (\planetradius/4);
\node at (\centerx,1.07*\centery-0.05,\centerz) {\scriptsize O};
\draw[load,color=red] (\centerx,\centery,\centerz) -- ++ (0,1.4*\planetradius,0) node[above]{\scriptsize $\bm{\Omega}$};
\draw[dotted] (\centerx+0.55*\planetradius/1.4142,\centery+0.55*\planetradius/1.4142,\centerz) -- (-\cubex,0,-\cubez);
\draw[dotted] (\centerx+0.55*\planetradius/1.4142,\centery+0.55*\planetradius/1.4142,\centerz) -- (-\cubex,0,0);
\draw[dotted] (\centerx+0.55*\planetradius/1.4142,\centery+0.55*\planetradius/1.4142,\centerz) -- (-\cubex,-\cubey,-\cubez);
\draw[dotted] (\centerx+0.55*\planetradius/1.4142,\centery+0.55*\planetradius/1.4142,\centerz) -- (-\cubex,-\cubey,0);
\draw[dashed] (\centerx,\centery,\centerz) -- ++ (0.55*\planetradius/1.4142,0.55*\planetradius/1.4142,0);
\draw[axis] (\centerx+0.55*\planetradius/1.4142,\centery+0.55*\planetradius/1.4142,\centerz) -- ++ (\planetradius/2.4,\planetradius/2.4,0) node[above]{\scriptsize $~~z$};
\draw[axis] (\centerx+0.55*\planetradius/1.4142,\centery+0.55*\planetradius/1.4142,\centerz) -- ++ (-\planetradius/5.3,\planetradius/5.3,0);
\node at (\centerx+0.25,\centery+1.45,\centerz) {\scriptsize $y$};
\draw[axis] (\centerx+0.55*\planetradius/1.4142,\centery+0.55*\planetradius/1.4142,\centerz) -- ++ (\planetradius/5.3,0,-\planetradius/5.3) node[below]{\scriptsize $~~~x$};
\node at (\centerx+0.55*\planetradius/1.4142,\centery+0.55*\planetradius/1.4142-0.5,\centerz) {\scriptsize M};
\draw[axis,color=black] (\centerx,\centery+0.8*\planetradius,\centerz) arc (90:45:0.8*\planetradius);
\node at (\centerx+\planetradius/3,\centery+0.855*\planetradius,\centerz) {\scriptsize $\Theta$};
%
\draw[dashed] (-\cubex/2,-\cubey/2,-\cubez/2) -- ++ (\cubex/2,0,0);
\draw[dashed] (-\cubex/2,-\cubey/2,-\cubez/2) -- ++ (0,0,\cubez/2);
\draw[dashed] (-\cubex/2,-\cubey/2,-\cubez/2) -- ++ (0,-\cubey/2,0);
\draw[black] (0,0,-\cubez) -- ++(-\cubex,0,0) -- ++(0,-\cubey,0) -- ++(\cubex,0,0) -- cycle;
\draw[black] (-\cubex,0,0) -- ++(0,0,-\cubez) -- ++(0,-\cubey,0) -- ++(0,0,\cubez) -- cycle;
\draw[black] (0,-\cubey,0) -- ++(-\cubex,0,0) -- ++(0,0,-\cubez) -- ++(\cubex,0,0) -- cycle;
\draw[black,fill=fluorescentorange!60,opacity=0.6] (0,0,0) -- ++(-\cubex,0,0) -- ++(0,-\cubey,0) -- ++(\cubex,0,0) -- cycle;
\draw[black,fill=fluorescentorange!60,opacity=0.6] (0,0,0) -- ++(0,0,-\cubez) -- ++(0,-\cubey,0) -- ++(0,0,\cubez) -- cycle;
\draw[black,fill=fluorescentorange!60,opacity=0.6] (0,0,0) -- ++(-\cubex,0,0) -- ++(0,0,-\cubez) -- ++(\cubex,0,0) -- cycle;
\draw[load,color=blue] (-0.9*\cubex/3,-\cubey/2,-\cubez/2) -- ++ (0,-0.85*\cubey/3,0) node[below]{\scriptsize $\bm{g}$};
\draw[black,fill=red!80,opacity=0.6] (0,-\cubey/2,0) -- ++(-\cubex,0,0) -- ++(0,0,-\cubez) -- ++(\cubex,0,0) -- cycle;
\draw[load,color=red] (-3.5*\cubex/5,-\cubey,-\cubez/2) -- ++ (-0.55*\cubex,1.2*\cubey,0) node[above]{\scriptsize $\bm{\Omega}$};
\draw[axis] (-\cubex/2,-\cubey/2,-\cubez/2) -- ++ (-1.25*0.7071*\cubex/2,0,-1.3*0.7071*\cubez/2) node[above]{\scriptsize $\chi$};
\draw[axis,color=black] (-\cubex/2,-\cubey/2,-4*\cubez/5) arc (84:103.18:4);
\node at (-2.12*\cubex/3,-0.98*\cubey/2,-4.5*\cubez/5) {\scriptsize $\alpha$};
\draw[axis] (-\cubex/2,-\cubey/2,-\cubez/2) -- ++ (-\cubex,0,0) node[left] {\scriptsize $y$};
\draw[axis] (-\cubex/2,-\cubey/2,-\cubez/2) -- ++ (0,0,-\cubez) node[above right] {\scriptsize $x$};
\draw[axis] (-\cubex/2,-\cubey/2,-\cubez/2) node[below right]{\scriptsize M} -- ++ (0,0.8*\cubey,0) node[above] {\scriptsize $z$};
\draw[axis,color=black] (-\cubex/2,-0.9*\cubey/4,-\cubez/2) arc (90:109.15:\cubey);
\node at (-3*\cubex/4,-0.95*\cubey/5,-\cubez/2) {\scriptsize $\Theta$};
\end{tikzpicture}
\begin{tikzpicture}[scale=0.804]
\node[color=white] at (0,6) {.};
\pgfmathsetmacro{\lag}{0.01}
\node[anchor=south west,inner sep=0] at (0,0)
    {\includegraphics[width=0.95\linewidth]{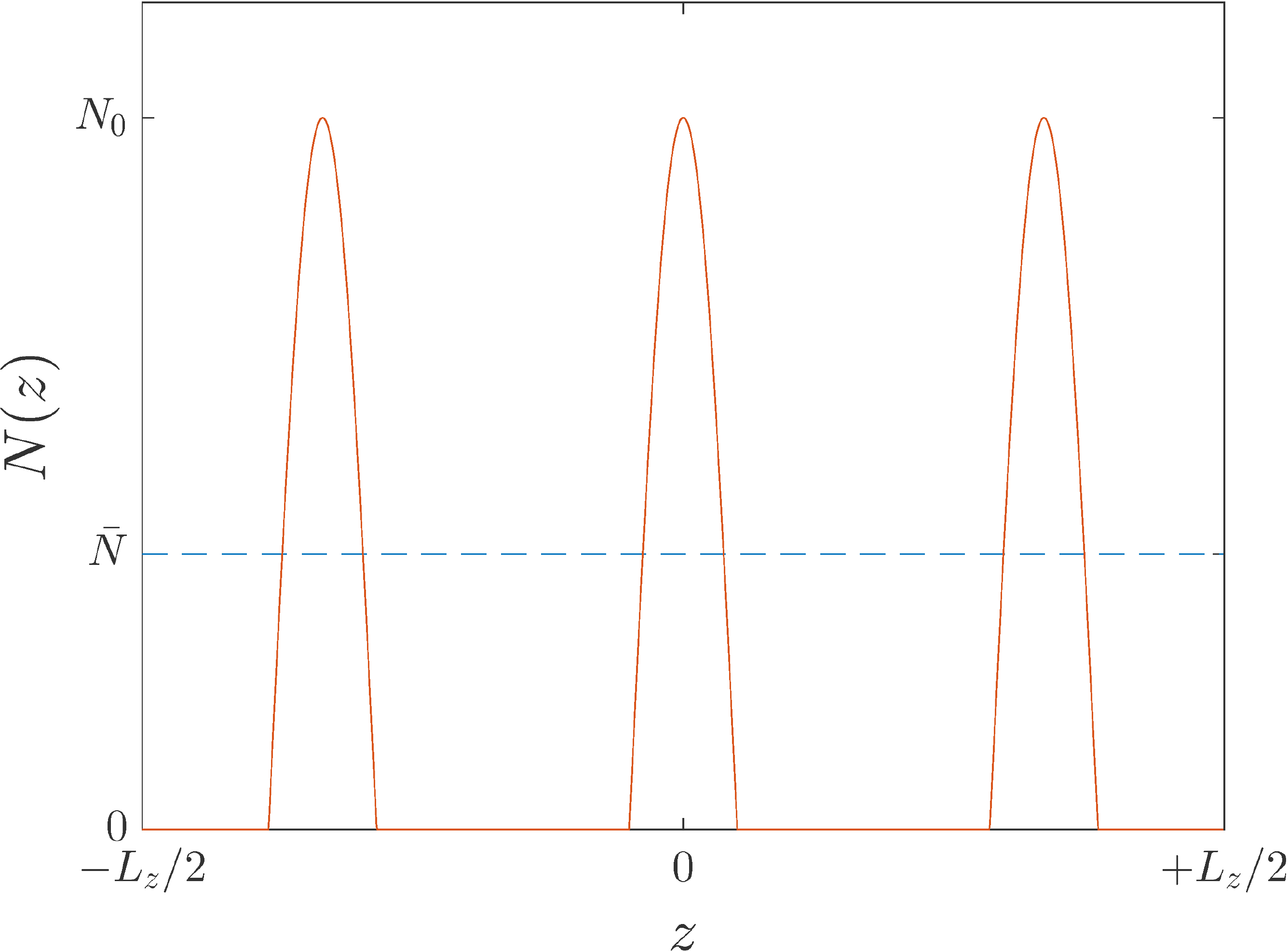}};
\draw[densely dotted,color=black!80] (0.8475,4.23+0.73) -- ++ (1.08,0);
\draw[densely dotted,color=black!80] (3.73,0.73) -- ++ (0,4.3);
\node[rotate=90] at (4.05,2.5) {\tiny Stably stratified interface};
\draw[densely dotted,color=black!80] (4.4,0.73) -- ++ (0,4.3);
\node[rotate=90] at (5.14,2) {\tiny Convective layer};
\draw[densely dotted,color=black!80] (5.875,0.73) -- ++ (0,4.3);
\draw[densely dotted,color=black!80] (5.875+4.4-3.73,0.73) -- ++ (0,4.3);
\draw[<->,color=black!80] (3.73+\lag,4.3+0.73) -- node[above]{\tiny $l$} (4.4-\lag,4.3+0.73);
\draw[<->,color=black!80] (4.4+\lag,4.3+0.73) -- node[above]{\tiny $d$} (5.875-\lag,4.3+0.73);
\draw[<->,color=black!80] (5.875+\lag,4.3+0.73) -- node[above]{\tiny $l$} (5.875+4.4-3.73-\lag,4.3+0.73);
\draw[<->,black!80] (5.875+4.4-3.73+\lag,4.3+0.73) -- node[above]{\tiny $d/2$} (5.875+4.4-3.73+0.7475-\lag,4.3+0.73);
\end{tikzpicture}%
\end{subfigure}
\hfill
\begin{subfigure}{0.55\textwidth}
\centering
\includegraphics[width=\linewidth]{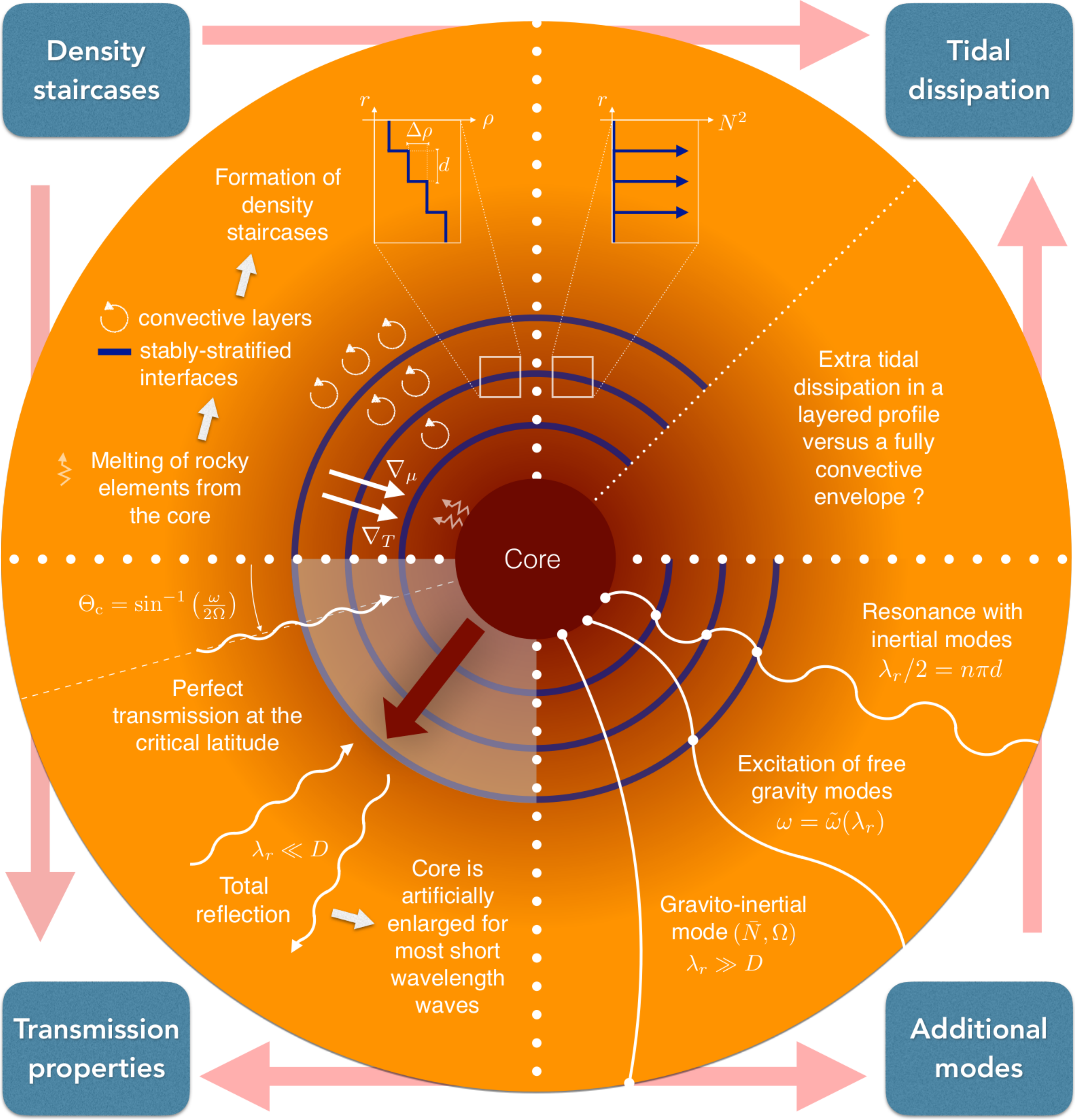}
\end{subfigure}
\caption{\textbf{Top-left:} our local Cartesian model, centred on a point M of a giant planet envelope at a colatitude $\Theta$. \textbf{Bottom-left:} example of buoyancy frequency profile containing three stably stratified layers. The thickness of the principal convective layers and stably stratified interfaces are $d$ and $l$, respectively. \textbf{Right:} summary sketch illustrating the physical context (top-right quarter), the transmission properties found in \cite{ABM2017} (bottom-left quarter), the additional modes {enabled} by a layered structure (bottom-right quarter), and the open question of whether tidal dissipation can be enhanced in a layered profile (top-right quarter).}
\label{fig:model}
\end{figure}
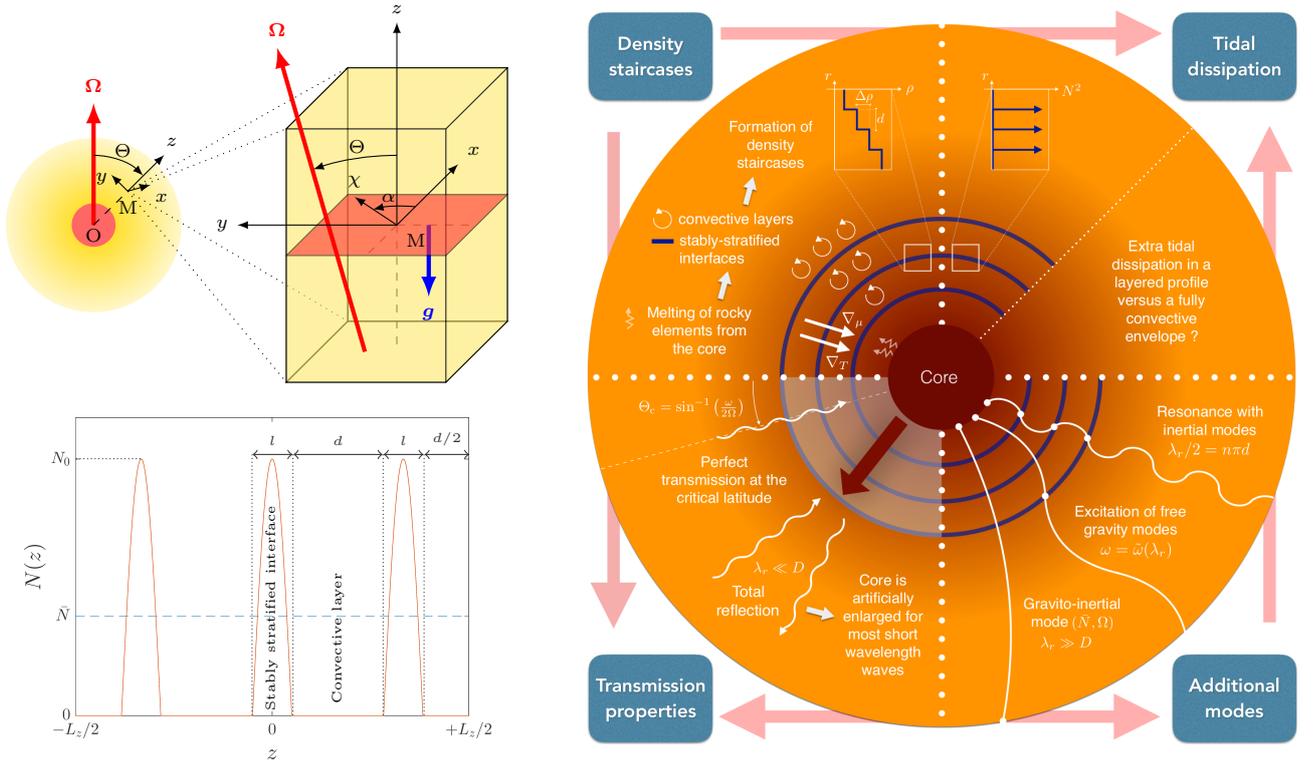

{To simplify our initial study, and because tidal waves typically have very short wavelengths,} we carry out our {analysis} in a local {Cartesian} box (denoted by $\mathcal{V}$) centered on a given point M of the {fluid} envelope, {with volume} $V$ (see top left panel of Fig. \ref{fig:model}). This allows us to study the local properties of the propagation and dissipation of internal waves in a region of layered semi-convection. Such an approach does not aim to give quantitative prediction{s} of the rates of tidal dissipation, but instead {can allow us to understand in detail the relevant physics}. In addition, it allows a {quantitative} comparison on the rates of tidal dissipation in a local region of layered semi-convection {versus} a fully convective medium. We take into account rotation through the Coriolis acceleration, the rotation vector having components $2\bm{\Omega}=2\Omega(0,\sin\Theta,\cos\Theta)$ in our local coordinate system, $\Theta$ standing for the colatitude and $\Omega$ the rotation rate.

The region of layered semi-convection is modelled by a buoyancy frequency profile, $N(z)$, that is zero in the convective regions and {positive} in stably stratified interfaces, as shown on the bottom left panel of Fig. \ref{fig:model}. The mean stratification {in} the vertical direction is $\bar{N}$. Solutions are taken to be periodic in both time and horizontal space coordinate, defined by $\chi=x\cos\alpha + y\sin\alpha$ (see top left panel of Fig. \ref{fig:model}). We also assume periodic boundary conditions in the vertical direction, which is equivalent to studying a small portion of a more vertically extended staircase. 

We study the linear propagation of {gravito-inertial waves (GIWs)} under the Boussinesq approximation subject to dissipative processes, namely viscosity and thermal diffusion, and including an external forcing $\bm{F} = (F_x,F_y,F_z)${, meant to mimic tidal forcing}. We refer the reader to \citet[hereafter ABM17]{ABM2017} for the system of equations composed of the momentum, continuity and heat transport equations. The buoyancy, pressure and velocity perturbations associated to GIWs are $b$, $p$ and $\bm{u}$, respectively.
The energy balance {satisfies} the following equation,
\begin{equation}
\frac{\text{d}\bar{E}}{\text{d}t} = -\frac{1}{V}\int_S \bm{\Pi}\cdot\text{d}\bm{S} + \bar{D}_{\text{visc}} + \bar{D}_{\text{ther}} + \bar{I},
\label{eq:energy_balance}
\end{equation}
where $\bar{E}$ is the total (pseudo-)energy of the wave, sum of kinetic and potential energies (averaged over the box), and $\bm{\Pi} = p\bm{u}$ is the flux density (flux per unit area) of energy. Then the energy dissipated by viscosity and thermal diffusion have the following expressions,
\begin{align}
\bar{D}_{\text{visc}} &= \frac{1}{V}\int_{\mathcal{V}} \rho_0\left( \nu \bm{u} \cdot \nabla^2 \bm{u} \right)\,\text{d}\mathcal{V},\\
\bar{D}_{\text{ther}} &= \left\{
\begin{array}{ccc}
\displaystyle \frac{1}{V}\int_{\mathcal{V}} \rho_0 \left(\frac{\kappa}{N^2}b \nabla^2b\right)\,\text{d}\mathcal{V} ~ & \text{if} & ~ N^2 \ne 0,\\[2.5mm]
0 ~ & \text{if} & ~ N^2 = 0,
\end{array}
\right.
\end{align}
respectively, while the energy injected by the forcing is $\bar{I} = \left(\int_{\mathcal{V}} \rho_0(\bm{u}\cdot\bm{F})\,\text{d}\mathcal{V}\right)/V.$ In the equations above, $\nu$ and $\kappa$ are the kinematic viscosity and thermal diffusivity, respectively. To quantify the relative importance of diffusive processes, we will use the Ekman number and its equivalent for thermal diffusion,
\begin{equation}
\text{E} = \frac{\nu}{2\Omega L_z^2} ~~~ \text{and} ~~~ K = \frac{\kappa}{2\Omega L_z^2},
\label{eq:EkmanThermal}
\end{equation}
respectively, where $L_z$ is the size of the box in the vertical direction. {These quantify the ratio of rotational to viscous (or thermal diffusion) timescales.}

Our goal is to calculate numerically\footnote{This is done using a Fourier collocation method \citep[see][]{Boyd2000}.} the average rate{s} of viscous and thermal dissipation in the box, $\bar{D}_{\text{visc}}$ and $\bar{D}_{\text{ther}}$, respectively, and the average rate of total dissipation in the box, $\bar{D} = \bar{D}_{\text{visc}} + \bar{D}_{\text{ther}}$.

\section{A first study of tidal dissipation in layered semi-convection}

\subsection{Propagation of tidal waves}
In \citetalias{ABM2017}, we studied in detail the effect of layered semi-convection upon the propagation of internal waves{. The} main properties are summarised and illustrated in the bottom-left quarter of the rightmost panel of Fig. \ref{fig:model}.

In particular, we found that an internal wave incident on a density staircase from above, will be reflected at the top unless it has a sufficiently large wavelength or is resonant with a free mode of the staircase. Thus, for most waves, the core of giant planets can be artificially enlarged by the presence of a region of layered semi-convection surrounding it. Moreover, we found that an inertial wave with frequency $\omega$ is perfectly transmitted at the location called the critical latitude, defined by $\theta_{\text{c}}=\sin\left(\omega/2\Omega\right)$. Finally, we were able to physically identify some specific modes that were perfectly transmitted through the staircase region. In what follows, we show that these modes correspond to resonant modes of the staircase, for which tidal dissipation is significantly enhanced.

\subsection{Dissipation rates in a layered profile}
\begin{figure}
\begin{tikzpicture}[scale=0.925]
\node[anchor=south west,inner sep=0] at (0,0)
    {\includegraphics[width=\textwidth]{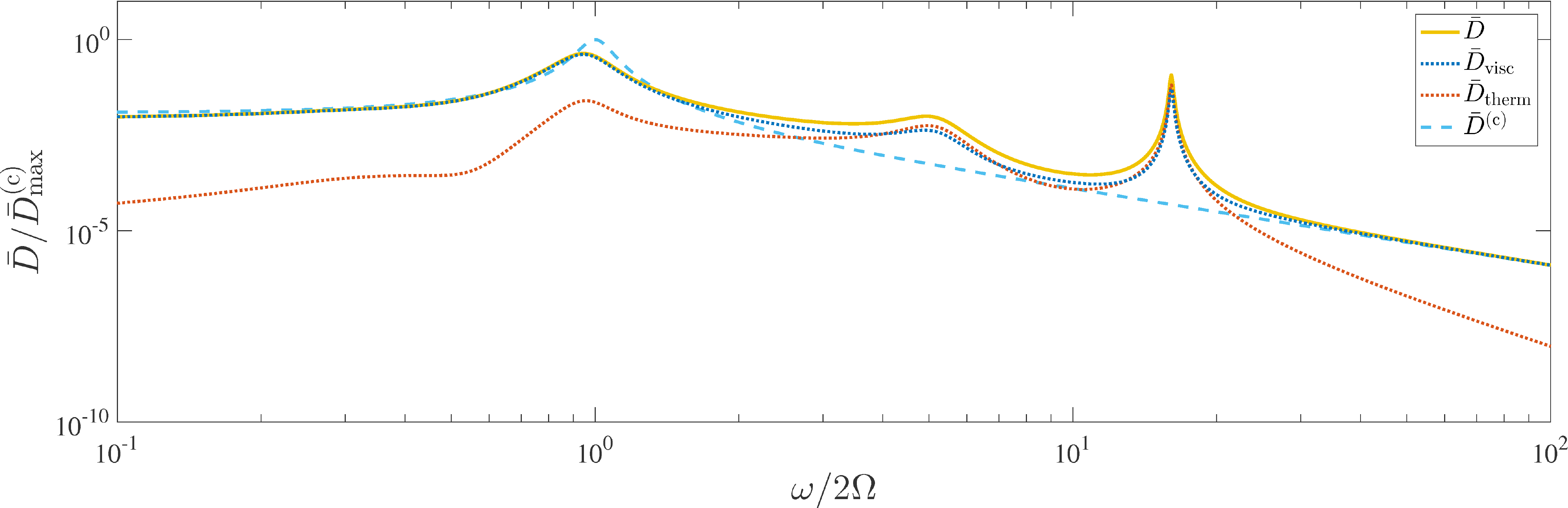}};
\node at (2.5,5.5) {\large $\text{E}=10^{-3}$};
\fill [color=red,opacity=0.08]
       (1.4,1.025) --++ (5.6,0) --++ (0,4.92) --++ (-5.6,0) -- cycle;
\fill [color=purple,opacity=0.08]
       (7,1.025) --++ (5.6,0) --++ (0,4.92) --++ (-5.6,0) -- cycle;
\fill [color=blue,opacity=0.08]
       (12.6,1.025) --++ (5.6,0) --++ (0,4.92) --++ (-5.6,0) -- cycle; 
\end{tikzpicture}
\begin{tikzpicture}[scale=0.925]
\node[anchor=south west,inner sep=0] at (0,0)
    {\includegraphics[width=\textwidth]{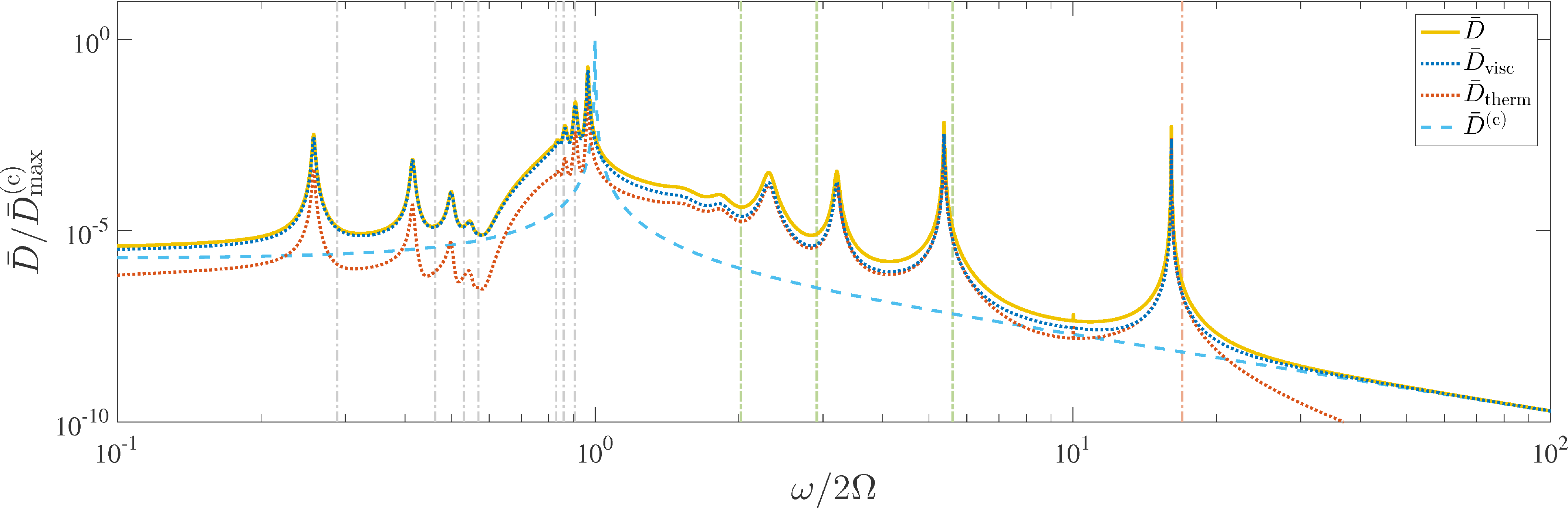}};
\node at (2.5,5.5) {\large $\text{E}=10^{-5}$};
\end{tikzpicture}
\caption{{Normalised dissipation as a function of tidal frequency} for different Ekman numbers, $\text{E}=K=10^{-3}$ (top panel) and $10^{-5}$ (bottom panel), {for} an aspect ratio $\varepsilon\equiv l/d=0.2$, and a colatitude $\Theta=\uppi/4$. The total dissipation is represented by the orange solid line, and its viscous and thermal contributions are represented by the dotted blue and red lines, respectively. The dashed light blue line represents the average dissipation spectrum for a fully convective box, $\bar{D}^{(\text{c})}$. For each panel, the quantity represented is $\bar{D}/\bar{D}^{(\text{c})}_{\text{max}}$, where $\bar{D}^{(\text{c})}_{\text{max}} \equiv \max_{\omega}\bar{D}^{(\text{c})}$, as a function of the normalised forcing frequency $\omega/2\Omega$. In the top panel, the pink, purple and blue regions correspond to resonances with inertial, super-inertial gravito-inertial and gravity modes, respectively. In the bottom panel, the vertical dashed-dotted lines indicate the position of the characteristic frequencies; namely, from left to right: resonance with short wavelength inertial modes (in grey), resonance with short wavelength gravito-inertial modes (in light green), and resonance with a free gravity mode of the staircase (in light red).}
\label{fig:spectra}
\end{figure}
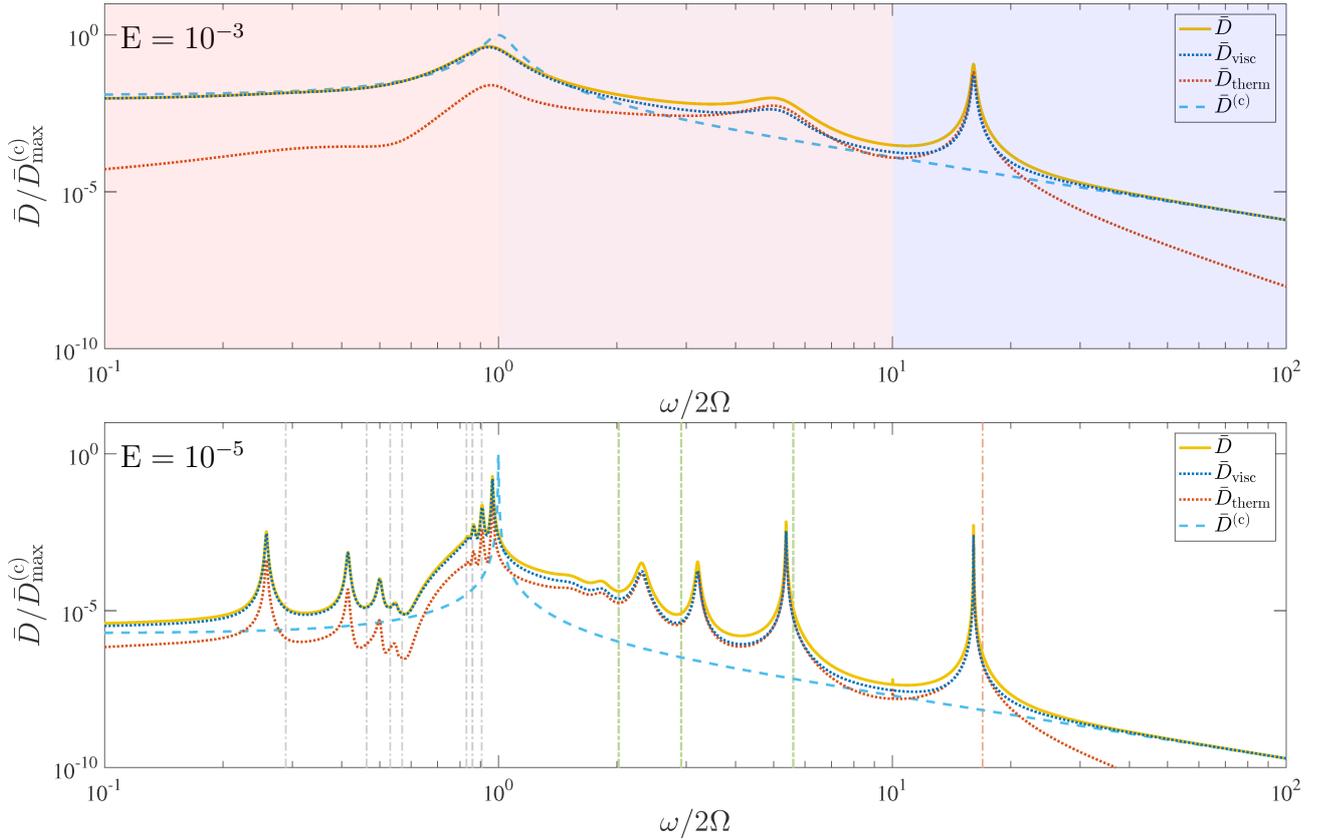

Here, we focus on the case {with} one stably stratified interface in the middle of our periodic box. The dissipation rates are plotted as a function of tidal forcing frequency on Fig. \ref{fig:spectra}, with the different panels corresponding to varying the diffusivity coefficients such that $\text{E} = K = 10^{-3}$ and $10^{-5}$ in the top and bottom panels, respectively. For each panel, the total dissipation rate, $\bar{D}$, is represented by the solid orange line, while its viscous and thermal contributions, $\bar{D}_{\text{visc}}$ and $\bar{D}_{\text{ther}}$, are represented by the dotted blue and red lines, respectively. For comparison, {we plot the corresponding quantities} in a fully convective medium, $\bar{D}^{(\text{c})}$ by the dashed light blue line. All dissipation rates are normalised by $\bar{D}^{(\text{c})}_{\text{max}} \equiv \max_{\omega}\bar{D}^{(\text{c})}$.

In agreement with \cite{OgilvieLin2004} or \cite{AuclairDesrotourEtal2015}, the resonant peaks are more numerous and narrower when the viscosity (and thermal diffusivity) is decreased to reach {smaller} Ekman numbers {that are more} relevant to planetary or stellar interiors. However, while our choice of parameters would give only one resonant peak in a uniformly stably stratified or fully convective medium -- centred on $\omega/2\Omega = 1$ in the latter case (see Fig. \ref{fig:spectra}) -- we clearly see here that the layered structure introduces new resonances. Focusing on the bottom panel, corresponding to the more relevant case with lower diffusivities, we clearly see that the total dissipation rate is higher in the layered case (orange curve) -- in particular in the sub-inertial range ($\omega < 2\Omega$) relevant to tidal forcing -- except near the Coriolis frequency ($\omega \sim 2\Omega$). This increase of the dissipation is allowed by the presence of new resonant peaks. Those additional resonances are broadly distributed over the frequency spectrum. Some correspond to resonances with inertial modes, corresponding to frequencies $\omega \lesssim 2\Omega$ (pink region on top panel of Fig. \ref{fig:spectra}); some with super-inertial gravito-inertial modes, corresponding to frequencies $2\Omega \lesssim \omega \lesssim \bar{N}$ (purple region on top panel of Fig. \ref{fig:spectra}); and finally some with gravity modes, corresponding to frequencies $\bar{N} \lesssim \omega \lesssim N_0$ (blue region on top panel of Fig. \ref{fig:spectra}).

\paragraph{Resonance with short wavelength inertial modes.}
In the inertial regime ($\omega<2\Omega$), a succession of resonant modes appear as we decrease the viscosity. We found that those correspond to inertial modes with vertical semi-wavelength{s} that fit{s} inside the convective layer. The grey dashed lines on the bottom panel of Fig. \ref{fig:spectra} thus correspond to frequencies that obey the relation $\lambda_z^{\text{(c)}}/2 = nd$ or equivalently $k_z^{(\text{c})}(\omega)d = n\uppi$, for different integers $n$. We recall that $d$ is the vertical extent of the convective region (see bottom left panel of Fig. \ref{fig:model}), $\lambda_z$ is the vertical wavelength and $k_z$ is the vertical wave number. In order to draw the vertical lines on Fig. \ref{fig:spectra}, we used the vertical wave number in the adiabatic limit, which can be obtained from the dispersion relation of pure inertial waves,
\begin{equation}
\omega^2 = \frac{(2\bm{\Omega}\cdot\bm{k})^2}{k_{\perp}^2+k_z^{(\text{c}) 2}},
\label{eq:disprelIW}
\end{equation}
where $\bm{k} = k_{\perp}\hat{\bm{e}}_{\chi} + k_z^{(\text{c})} \hat{\bm{e}}_z$. We have used the label $^{\text{(c)}}$ above to stress that this is the expression of the vertical wave number in a convective medium.

The discrepancy between the prediction and the actual position of those resonant modes can be partly explained twofold. First, the expression of the vertical wave number that was used corresponds to the adiabatic case. Second, the stably stratified interface in the middle of the box has a non-negligible vertical extent in which pure inertial waves become influenced by buoyancy. This differs from the simplified model that {we primarily studied} in \citetalias{ABM2017}, in which stably stratified interfaces were infinitesimally small.

\paragraph{Resonance with short wavelength super-inertial gravito-inertial modes.}
Based on the same idea, we looked for modes in the gravito-inertial regime {with} vertical semi-wavelengths {that} fit inside the vertical extent of the {thin} stably stratified layer, in which they are propagative. On the bottom panel of Fig. \ref{fig:spectra}, the dotted-dashed light green vertical lines correspond to frequencies such that $k_z(\omega)d = n\uppi$ for three different integers $n$, {satisfying} the dispersion relation of gravito-inertial waves,
\begin{equation}
\omega^2 = N_{\text{e}}^2\frac{k_{\perp}^2}{k_{\perp}^2+k_z^2} + \frac{(2\bm{\Omega}\cdot \bm{k})^2}{k_{\perp}^2+k_z^2}.
\label{eq:GIWs}
\end{equation}
{Here} $k_z$ is the vertical wave number in a stably stratified medium, characterised by the constant buoyancy frequency $N_{\text{e}}$. We note that these waves are evanescent in the convective layer surrounding the stably stratified interface \citep[see also][for an extensive discussion]{MathisNeinerTranminh2014}.

The buoyancy frequency {is not} uniform in the stably stratified region (see bottom left panel of Fig. \ref{fig:model}), {therefore} $N_{\text{e}}$ was somewhat tuned so that it would match the corresponding peaks as close as possible. Its value was found to be consistent with the requirement that this effective buoyancy frequency, $N_{\text{e}}$, should lie in the range $\bar{N} < N_{\text{e}} < \max_z N(z)$. However, following this procedure, we do not expect the vertical lines to match perfectly the position of the resonant peaks, {but the agreement is sufficient to validate our interpretation.}

\paragraph{Resonance with the free gravity mode of the staircase.}
In \citetalias{ABM2017}, {we derived the dispersion relation for the free modes of the staircase, by extending} \cite{BelyaevQuataertFuller2015}. When looking for gravity modes, {we can neglect} rotation {so that the} dispersion relation reduces to
\begin{equation}
\omega^2 = \bar{N}^2 \left(
\frac{k_{\perp}d}{2\coth (k_{\perp}d)-2\cos\theta\csch (k_{\perp}d)}
\right),
\label{eq:BQFdisprel_finite_NoRotation}
\end{equation}
where $\cos\theta$ is a root of a polynomial \citepalias[see][]{ABM2017}. {This predicts} the free modes of a layered structure {in} which stably stratified interfaces are modelled as discontinuous jumps.
The frequency given by the equation above is displayed as the dotted-dashed light red vertical line on the bottom panel of Fig. \ref{fig:spectra}. Its position matches rather well the position of the rightmost resonant peak, though we do not expect a perfect match because we have neglected the effect{s} of rotation, viscous and thermal {diffusions, and the interfaces are not infinitesimally thin}.

\section{Conclusions}
We computed the dissipation rates in a region of layered semi-convection as a function of tidal frequency {using a local Cartesian model}. We found that the dissipation of tidal waves can be significantly enhanced compared to a fully convective medium {in the sub-inertial frequency range, which is typically the most relevant for tidal forcing in giant planet systems}. In addition, such a {density} structure leads to new dissipation peaks {corresponding with additional resonances}, for which the dissipation can be increased by several orders of magnitude. We {tentatively} conclude that layered semi-convection is a possible candidate to explain {the} high tidal dissipation rates observed by \cite{LaineyEtal2009,LaineyEtal2012,LaineyEtal2017} in Jupiter and Saturn. {Further work is required to explore and confirm the influence of layered semi-convection on tidal dissipation in global models.} In a near future, other mechanisms such as differential rotation and magnetic fields should also be taken into account \citep{BaruteauRieutord2013,BarkerLithwick2013,Wei2016,LinOgilvie2017}.

\begin{acknowledgements}
QA and SM acknowledge funding by the European Research Council through ERC SPIRE grant 647383. AJB was supported by the Leverhulme Trust through the award of an Early Career Fellowship.
\end{acknowledgements}

\bibliographystyle{aa}  
\bibliography{andre} 

\end{document}